\documentclass[lettersize,journal]{IEEEtran}
\usepackage{amsmath,amsfonts}
\usepackage{algorithmic}
\usepackage{algorithm}
\usepackage{array}
\usepackage[caption=false,font=normalsize,labelfont=sf,textfont=sf]{subfig}
\usepackage{textcomp}
\usepackage{stfloats}
\usepackage{url}
\usepackage{verbatim}
\usepackage{graphicx}
\usepackage{cite}
\usepackage{lipsum}
\usepackage{xcolor}
\usepackage[nolist]{acronym}
\usepackage{subfiles}
\usepackage{siunitx}
\usepackage{tikz}
\usetikzlibrary{arrows.meta}
\usetikzlibrary{patterns}
\usetikzlibrary{calc}
\usetikzlibrary{spy}

\usepackage[utf8]{inputenc}
\usepackage{pgfplots}
\DeclareUnicodeCharacter{2212}{−}
\usepgfplotslibrary{groupplots,dateplot}
\usetikzlibrary{patterns,shapes.arrows}
\pgfplotsset{compat=newest}

\usepackage{amsmath}

\begin{document}

\title{Joint Communication and Sensing: 5G NR Compliant Ranging Using the Sounding Reference Signal}

\author{Michael Hofstadler,~\IEEEmembership{Student Member,~IEEE,}
	Reinhard Feger,
	Andreas Springer,~\IEEEmembership{Member,~IEEE,}
	Andreas Stelzer,~\IEEEmembership{Member,~IEEE,}
	Harald Pretl,~\IEEEmembership{Senior Member,~IEEE}
\thanks{Manuscript received Month Day, Year; revised Month Day, Year.}
\thanks{Michael Hofstadler, Reinhard Feger, Andreas Springer and Andreas Stelzer are with the Institute for Communications Engineering and RF-Systems, Johannes Kepler University, Linz 4040, Austria (e-mail: \hbox{michael.hofstadler@jku.at}; \hbox{reinhard.feger@jku.at}; \hbox{andreas.springer@jku.at}; \hbox{andreas.stelzer@jku.at}). Harald Pretl is with the Institute for Integrated Circuits, Johannes Kepler University, Linz 4040, Austria (e-mail: \hbox{harald.pretl@jku.at})}}
%

\markboth{Journal of \LaTeX\ Class Files,~Vol.~14, No.~8, August~2021}%
{Shell \MakeLowercase{\textit{et al.}}: A Sample Article Using IEEEtran.cls for IEEE Journals}

\IEEEpubid{0000--0000/00\$00.00~\copyright~2021 IEEE}

\maketitle

\begin{abstract}


In this work, a proof of concept for 5G-compliant user-equipment side sensing is presented. It is based on orthogonal frequency division multiplexing radar-based ranging which is realized in this work by using the sounding reference signal from the 5G New Radio standard. The signal configuration and thus the corresponding waveform is generated in compliance with the existing \acl{3gpp} standard for 5G. It is an uplink physical signal and is originally intended, amongst others, for channel estimation. The used model is introduced, followed by the sounding reference signal. This leads to a first proof of concept by presenting simulation and measurement results. We show, that a range estimation error in the order of centimeters is achievable.
\end{abstract}


\section{Introduction}
\IEEEPARstart{W}{ith} the consecutive release of standards for broadband cellular networks, like 5G and its predecessor, both new challenges and opportunities arise. One prominent example for this is the introduction of the so-called \ac{fr2} in 5G ranging from $24.25$ to \SI{52.6}{\giga \hertz} \cite{3gpp.38.101-2}. The increased bandwidths available in \ac{fr2} provide a promising  base for new opportunities in terms of sensing, with an even increasing trend towards 6G. This is also reflected by the popularity gaining research area of \acf{jcas} \cite{9694609}, also known as \ac{jrc}, which can be seen, e.g., in the appearance of the first two editions of the IEEE JC\&S Symposium \cite{jcns}. Although research for bringing communication and sensing together exist for a while already \cite{858893}, a large number of  new research efforts within the different areas of \ac{jcas}  can be found nowadays. %
The majority of them address an area of vehicular side \cite{8246850},  or network side \ac{jcas}, addressing a variety of scopes, from more general, network-wide concepts \cite{9354629},  to the analysis of \ac{jcas} performance within a \ac{bs}   \cite{9201077} . %
Less research effort, on the other hand, goes to \ac{ue} side \ac{jcas}. Reference \cite{9115568} investigates \ac{ue} side \ac{jcas} for grid-based indoor mapping, utilizing a \ac{nr} uplink transmit signal.


In this work, we present a \ac{ue} side \ac{jcas} approach. We focus on embedding ranging into an existing 5G communication scenario. Thus we need to be fully compliant with the 5G \ac{nr} standard as released by the \ac{3gpp} \cite{3gpp.38.101-2,3gpp.38.211}. Therefore, it is obvious that we perform ranging based on an \ac{ofdm}  radar, which is already well established \cite{9442375}. Having the \ac{ue} side in scope, the used signal has to be an \ac{ul} signal. Here we propose to use the \ac{srs}. This signal was chosen for a number of reasons, which will be illustrated in the following section.

\IEEEpubidadjcol

\section{\ac{ofdm}-Based Ranging and  the Sounding Reference Signal}
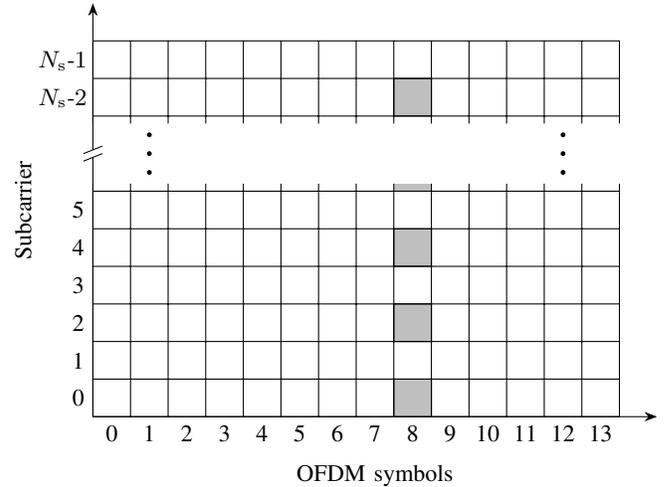
\begin{figure}[!t]
	\centering
	\begin{tikzpicture}[scale=0.5]
	\begin{small}
	
	
	\draw[-Stealth] (0,0) -- (15,0) node[midway, below=15pt] {OFDM symbols};
	\draw[-Stealth] (0,0) -- (0,11) node[rotate=90, midway, above=20pt] {Subcarrier};
	
	\coordinate (break) at (0,7);
	\draw[white, line width=1pt] ($ (break) + (0,-0.1) $) -- ($ (break) + (0,0.1) $);
	\draw ($ (break) + (-0.25,0) $) -- ($ (break) + (0.25,0.2) $);
	\draw ($ (break) + (-0.25,-0.2) $) -- ($ (break) + (0.25,0) $);

	\foreach \x in {0,...,13}
	\draw ($(\x,0) + (0.5,0)$) node[below] {\x};
	
	\foreach \y in {0,...,5}
	\draw ($(0,\y) + (0,0.5)$)  node[left] {\y};
	
	\draw (2pt,8) node[above left] {$N_{\mathrm{s}}$-$2$};
	\draw (2pt,9) node[above left] {$N_{\mathrm{s}}$-$1$};
	
	
	\draw[fill=lightgray]  (8,0) rectangle ++(1,1);
	\draw[fill=lightgray]  (8,2) rectangle ++(1,1);
	\draw[fill=lightgray]  (8,4) rectangle ++(1,1);
	\draw[fill=lightgray, draw opacity=0,fill opacity=1]  (8,6) rectangle ++(1,0.2);
	\draw[fill=lightgray]  (8,8) rectangle ++(1,1);
	
	
	\draw[step=1] (0,0) grid (14,6.2);
	\draw[step=1] (0,7.8) grid (14,10);
	
	
	\coordinate (dots) at (1.5,6.5);
	\draw[fill=black] (dots) circle [radius=0.05];
	\draw[fill=black] ($ (dots) + (0,0.5) $) circle [radius=0.05];
	\draw[fill=black] ($ (dots) + (0,1) $) circle [radius=0.05];
	
	\draw[fill=black] ($ (dots) + (11,0) $) circle [radius=0.05];
	\draw[fill=black] ($ (dots) + (11,0.5) $) circle [radius=0.05];
	\draw[fill=black] ($ (dots) + (11,1) $) circle [radius=0.05];

	\end{small}
\end{tikzpicture}
	\caption{A slot of the radio frame from the resource grid. In the \ac{ofdm} symbol number eight, every second subcarrier is occupied by a complex-valued symbol, denoted by the gray filled squares.}
	\label{fig_1}
\end{figure}
The \ac{ue} is regularly communicating with a \ac{bs} when connected to the network. Even if no user data is actively transferred, the link partners, i.e. \ac{bs} and \ac{ue}, regularly send pilot signals, like control-, synchronization-, and sounding-signals. Ranging at the \ac{ue} can be performed if the reflections of \ac{ul} pilot signals stemming from objects in the vicinity are received and processed properly.

Before we elaborate on the chosen UL pilot signal we will introduce the 5G NR frame- and resource structure and the principle of \ac{ofdm}-based ranging.

\subsection{5G NR Frame- and Resource Structure}
The available time and frequency resources in an \ac{ofdm}-based mobile communication network are organized in a resource grid. This grid is a matrix of resource grid elements, where the rows $k$ represent the individual subcarriers, i.e. the frequency resources, whereas columns $l$ represent the \ac{ofdm} symbols according to the time resources. In the time domain, \ac{ofdm} symbols are grouped into slots and slots are grouped into subframes which are grouped into a radio frame. In this work only a single slot is taken into account, as
this is the smallest interval for consecutive \ac{ofdm} symbols.\footnote{Here, we do not consider mini-slot transmission, which can be used for ultra-reliable low-latency communication \cite{3gpp.38.912}.} The subcarrier spacing $\Delta f$ and thus the \ac{ofdm} symbol duration, the cyclic prefix length and the number of \ac{ofdm} symbols per slot are defined by the so-called numerology $\mu$ \cite{3gpp.38.101-2}. The subcarrier spacing $\Delta f$ is given by
\begin{equation}
	\label{deqn_1}
	\Delta f = 2^{\mu} \cdot 15  \, \mathrm{kHz}.
\end{equation}
For this work \hbox{$\mu=3$} is selected, which results in a subcarrier spacing of \hbox{$\Delta f =$ \SI{120}{kHz}}. For the chosen $\mu$ a slot comprises $14$ \ac{ofdm} symbols.

If a resource grid element is occupied, it is associated with a complex-valued \ac{ofdm} transmission symbol \hbox{$a_{k,l}$ with $|a_{k,l}| > 0$}. In all unoccupied resource grid elements $a_{k,l}=0$. Fig.~\ref{fig_1} shows the resource grid of one slot, where every second subcarrier of the \ac{ofdm} symbol number eight is occupied. All other resource grid elements are unoccupied.



\subsection{\ac{ofdm}-Based Ranging}
The complex-valued symbols for the resource grid elements are given by \hbox{$a_{k,l}=A_{k,l}\mathrm{e}^{\mathrm{j}\Psi_{k,l}}$}, where $A_{k,l} \in \mathbb{R}$ is the amplitude and $\Psi_{k,l}$ is the phase of the complex-valued symbol. The continuous-time \ac{ofdm} baseband signal $s_{l}(t)$ transmitted by the \ac{ue}, including the cyclic prefix, for the $l$th \ac{ofdm} symbol and $N_{\mathrm{s}}$ subcarriers is given by~\cite{3gpp.38.211}
\begin{equation}
	\label{deqn_2}
	s_{l}(t) = 
	\begin{cases} 
		\bar{s}_{l}(t) & 0 \leq t <  T_{\mathrm{s}}\\
		0 & \mathrm{otherwise}
	\end{cases}
\end{equation}
with
\begin{equation}
	\label{deqn_2b}
	\bar{s}_{l}(t) =\sum_{k=0}^{N_{\mathrm{s}}-1} a_{k,l} \mathrm{e}^{\mathrm{j} 2 \pi (k-N_{\mathrm{s}}/2) \Delta f (t-T_{\mathrm{cp}})} ,
\end{equation}%
where \hbox{$T_{\mathrm{s}} = T_{\mathrm{symb}} + T_{\mathrm{cp}}$} is the sum of the symbol time duration \hbox{$T_{\mathrm{symb}}=1/\Delta f$} and the cyclic prefix time duration $T_{\mathrm{cp}}$. For the sake of simplicity, compared to \cite{3gpp.38.211}, the subcarrier starting point $k_{\mathrm{0}}$ and the starting point in time $t_{\mathrm{start}}$ has been set to zero and the antenna port $p$ has been set to one. From (\ref{deqn_2}) we can compute the \ac{rf} signal $s(t)$ transmitted at the carrier frequency $f_{0}$ with the amplitude scaling factor $\beta$ as
\begin{equation}
	\label{deqn_3}
	s(t) = \operatorname{Re}\{  \beta \, s_{l}(t) \mathrm{e}^{\mathrm{j} 2 \pi  f_{0} (t-T_{\mathrm{cp}}   )} \}.
\end{equation}
%
%
We assume the transmitted signal in (\ref{deqn_3}) to be reflected by an object. Thus the signal $x(t)$ received by the \ac{ue} is a scaled (by $\alpha$) and delayed (by $\tau$, which depends on the distance between UE and object) version of the TX signal corrupted by noise $n(t)$.
\begin{equation}
	\label{deqn_4}
	x(t) = \alpha \, s(t-\tau) + n(t)
\end{equation}
After downconversion, cyclic prefix removal and demodulation, the $k$th subcarrier element of the $l$th \ac{ofdm} symbol of the baseband signal is given by
\begin{equation}
	\label{deqn_5}
	X_{l}[k]=\alpha \beta A_{k,l}\mathrm{e}^{\mathrm{j}(\Psi_{k,l} - 2 \pi \tau\{k \Delta f+f_{0}\})} + N_{k}^{\prime},
\end{equation}
where $N_{k}^{\prime}$ represents the processed noise.

The range $R$ between the \ac{ue} and the object is related to $\tau$ and the speed of light in the corresponding medium $c$ by \hbox{$R=c \tau / 2$}. To estimate $\tau$ and thus $R$, the baseband signal $X_{l}$ needs to be further processed to extract the range information.
%
%

To remove the individual phase and amplitude information coming from the complex-valued transmission \ac{ofdm} symbols $a_{k,l}$, an element-wise division is applied to arrive at  
\begin{equation}
	\label{deqn_7}
	\varepsilon_{l}[k]=\frac{X_{l}[k]}{a_{k,l}}=\alpha \beta \gamma \mathrm{e}^{  - \mathrm{j} 2 \pi \tau k \Delta f} + N_{k}.
\end{equation}
where \hbox{$\gamma=e^{- \mathrm{j} 2 \pi \tau f_{0}}$}, describes a phase offset  constant over $k$.

To calculate the complex-valued range profile $\tilde{\rho}[n]$, a \ac{fft} along the subcarrier index $k$ is performed with
\begin{equation}
	\label{deqn_8}
	\tilde{\rho}[n] = \sum_{k=0}^{N_{\mathrm{FFT}}-1} \varepsilon_{l}[k] \mathrm{e}^{\mathrm{j} 2 \pi k \Delta f \tau_{n}}.
\end{equation}
where $N_{\mathrm{FFT}}$ is the number of \ac{fft} points and \hbox{$\tau_{n}=n/(N_{\mathrm{FFT}}\Delta f)$} are the discretized time delay steps. According to the \ac{3gpp} standard an  \hbox{$N_{\mathrm{FFT}} = 4096$} point \ac{fft} has been used. The range profile is then given by \hbox{$\rho[n]=|\tilde{\rho}[n]|^{2}$}. The estimated time delay $\tau_{\hat{n}}$, according to the peak position \hbox{$\hat{n}=\arg \max_{n} \rho[n]$}, then represents the nearest neighbor to the true time delay of $x(t)$ and the corresponding range $R$.%
%

To increase the precision of the estimation, an interpolation based on the computationally efficient \ac{czt} can be performed \cite{1162132}. The domain used for this operation is limited to the vicinity around the peak from $\tau_{\hat{n}-1}$ to $\tau_{\hat{n}+1}$. This interpolated range profile is denoted as $\rho_{\mathrm{\textsc{czt}}}[m]$. With $N_{\mathrm{FFT}} $ interpolation points, this leads to a more precise discretization of \hbox{$\tau_{m}=2m/(N_{\mathrm{FFT}}^{2}\Delta f)$}. The now received peak position \hbox{$\hat{m}=\arg \max_{m} \rho_{\mathrm{\textsc{czt}}}[m]$} and the according  time delay $\tau_{\hat{m}}$ is finally used to estimate the range $\hat{R}$ to the object, which can be computed as
\begin{equation}
	\label{deqn_9}
	\hat{R} = \frac{\mathrm{c}_{0}}{2} \tau_{\hat{m}}.
\end{equation}

\subsection{The Sounding Reference Signal}
The \ac{srs} is used, amogst others, for channel estimation in the \ac{ul}. Therefore, it is also sent in situations when no user data is sent. It is based on a low \ac{papr} sequence. Depending on the length of the sequence it is a Zadoff-Chu sequence, or a numerically derived sequence with properties which fulfill the requirements of the standard similarly as the Zadoff-Chu sequence does \cite{3gpp.38.211}. Therefore, the \ac{srs} sequence is a \ac{cazac} sequence. The \ac{papr} characteristic is also beneficial for the transmit signal, as the \ac{papr} of the waveform is limited, and the requirements on the transceiver are lower, compared to an \ac{ofdm} transmit signal carrying a random bit stream.


The sequence is allocated  in the resource grid according to the comb number \hbox{$K_{\mathrm{TC}}  \in \{2,4,8\}$}. This number defines how every $K_{\mathrm{TC}}$th subcarrier is used for \ac{srs} sequence allocation. A resource grid occupied by an \ac{srs} sequence with \hbox{$K_{\mathrm{TC}}=2$} can be seen in Fig.~\ref{fig_1}. This also means, that even if the subcarrier spacing is \hbox{$\Delta f =$ \SI{120}{\kilo \hertz}}, the smallest frequency difference for a single \ac{srs} sequence is \hbox{$\operatorname{min}(K_{\mathrm{TC}}) \Delta f =$ \SI{240}{\kilo \hertz}}. The sequence can occupy \hbox{$N_{\mathrm{symb}}^{\mathrm{SRS}} \in \{1, 2, 4,8,10,12,14\}$}  consecutive \ac{ofdm} symbols per slot.

In terms of subcarriers, the number of resource elements which can be occupied ranges from 24 up to 1584 in a single-carrier configuration. The number of occupied subcarriers therefore also corresponds to the bandwidth of the signal.

The \ac{srs} is also used in the so-called \textit{beam mobility procedure}. As beamforming is part of 5G \ac{nr}, selecting the best beam for transmission is of high interest for both link partners. Within this \ac{ul} procedure, the \ac{bs} performs power measurements, while the \ac{ue} is transmitting the \ac{srs} on one or more beams \cite{kottkamp20195g}. Thus, the \ac{srs} could be used in a similar manner for \ac{ue}-side ranging.
\section{Experimental Verification Methods}
\begin{figure}[!t]
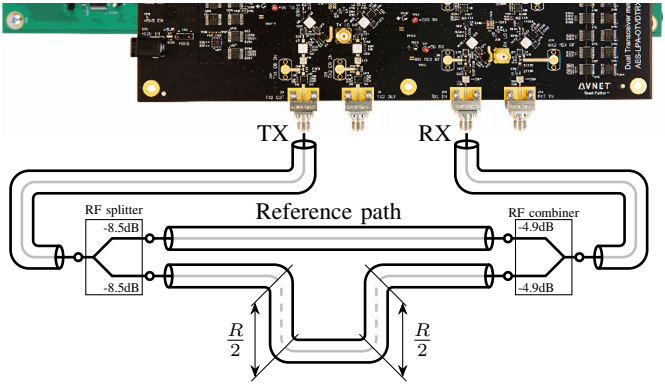

	\centering
	\subfile{./figures/measurement_setup}
	\caption{The measurement setup with a cable-based measurement scenario. The \ac{rf} network, based on coaxial components consist of two paths, connected by \ac{rf} splitters/combiners. The reference path represents the TX/RX leakage and a longer, second path represents the distance to a target of interest. The dashed lines represents the difference in length $R$ of the two paths.}
	\label{fig_2}
\end{figure}
 \begin{figure}[!t]
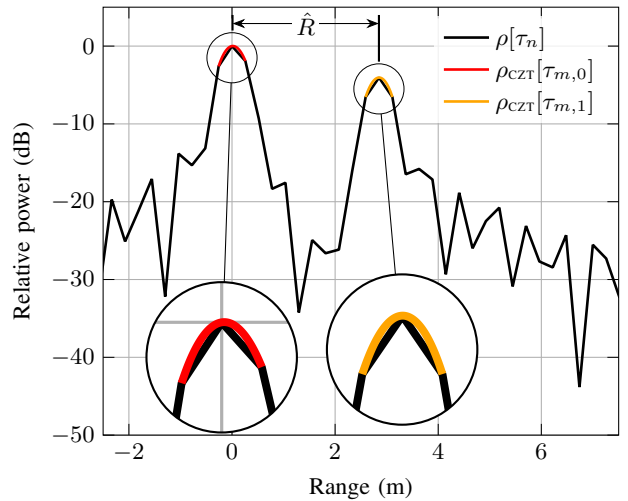

 	\centering
 	\subfile{./figures/range2}
 	\caption{The range profile $\rho[\tau_{n}]$, from the measurement results according to the setup with a path difference of $R=\SI{2.76}{\metre}$, normalized to the strongest peak. The \ac{czt} interpolation around the peak for the reference path $\rho_{\mathrm{\textsc{czt}}}[\tau_{m,0}]$ is shown in red, the one around the peak caused by the second longer path $\rho_{\mathrm{\textsc{czt}}}[\tau_{m,1}]$ in orange. The distance between the two peaks is the estimated range $\hat{R}$.}
 	\label{fig_3}
 \end{figure}
To verify the concept of ranging with a 5G \ac{nr} compliant signal, a software framework has been created. This framework has been used to create standard compliant carrier and \ac{srs} configurations, and to generate standard compliant waveforms \cite{matlab5g}. The framework has also been used for simulations. The simulated scenarios have been set to match the measurement setup, to allow comparison of the results. For the measurement setup, an \textit{AMD-Xilinx RFSoC Gen 3 Kit for mmWave} with a configured carrier frequency of \hbox{$f_{0} =$ \SI{25}{\giga \hertz}} has been used. The hardware has been controlled by the \textit{Avnet RFSoC Explorer} software.

As there is no synchronization between TX and RX in the default \ac{fpga} configuration, a reference target is introduced. The delay of the reference target was chosen to be shorter than the target of interest. This reference target is also motivated due to the fact, that in the \ac{ue}, TX and RX are also not precisely synchronized, but a TX-to-RX leakage signals is present, which can be exploited for synchronization purposes. The two different targets and therefore the according time delays are denoted as $\tau_{m,i}$, where \hbox{$i = \{0,1\}$} is the index of the two most predominant peaks in the range profile. The estimated time delays are the one from the reference path $\tau_{\hat{m},0}$ and the one representing the target distance of interest $\tau_{\hat{m},1}$. The difference \hbox{$\tau_{\hat{m},\delta} = \tau_{\hat{m},0}-\tau_{\hat{m},1}$} is then related to the distance between the reference target and the target of interest
\begin{equation}
	\label{deqn_9}
	\hat{R} = \frac{\mathrm{c}_{0}}{2} \tau_{\hat{m},\delta}.
\end{equation}
Fig.~\ref{fig_2} shows the measurement setup. The measurements have been performed with a cable-based measurement scenario. The \ac{rf} network, based on coaxial components, consists of two paths, connected by \ac{rf} splitters/combiners. The reference path represents the TX/RX leakage and the longer, second path represent the distance to a target of interest. The dashed lines represents the difference in length $R$ of the two paths. Therefore the relation between estimated range and time delay changes to
\begin{equation}
	\label{deqn_9}
	\hat{R} = v_{p} \tau_{\hat{m},\delta},
\end{equation}
where $v_{p}$ is the phase velocity in the coaxial lines. 

\begin{figure*}[!t]
	\centering
	\subfloat[]{
\begin{tikzpicture}
	\begin{small}

\definecolor{darkgray176}{RGB}{176,176,176}

\begin{axis}[
legend cell align={left},
legend style={fill opacity=0.8, draw opacity=1, text opacity=1, draw=none},
log basis y={10},
tick pos=both,
x grid style={darkgray176},
xlabel={Bandwidth (MHz)},
xmajorgrids,
xmin=-12.96, xmax=398.88,
xtick style={color=black},
xtick={-100,0,100,200,300,400},
xticklabels={
  \(\displaystyle {\ensuremath{-}100}\),
  \(\displaystyle {0}\),
  \(\displaystyle {100}\),
  \(\displaystyle {200}\),
  \(\displaystyle {300}\),
  \(\displaystyle {400}\)
},
y grid style={darkgray176},
ylabel={Estimation error (m)},
ymajorgrids,
ymin=0.0003, ymax=100,
ymode=log,
ytick style={color=black},
ytick={1e-05,0.0001,0.001,0.01,0.1,1,10,100,1000},
yticklabels={
  \(\displaystyle {10^{-5}}\),
  \(\displaystyle {10^{-4}}\),
  \(\displaystyle {10^{-3}}\),
  \(\displaystyle {10^{-2}}\),
  \(\displaystyle {10^{-1}}\),
  \(\displaystyle {10^{0}}\),
  \(\displaystyle {10^{1}}\),
  \(\displaystyle {10^{2}}\),
  \(\displaystyle {10^{3}}\)
}
]
\addplot [black, mark=square*, mark size=1.5, mark options={solid}, line width=1pt]
table {%
5.76 59.3639988710285
11.52 29.5726893287992
23.04 13.2009490247727
28.8 10.0052529576349
40.32 6.40794315392296
51.84 4.62745136936864
57.6 9.60610398289314
69.12 8.38126435830146
80.64 0.81091843557561
92.16 0.881545748846556
103.68 0.610807714641516
115.2 0.331715912200215
126.72 0.0708505131836557
138.24 0.11774213336408
149.76 0.206722422305662
161.28 0.140525137644925
172.8 0.182866950898203
184.32 0.301465145404988
195.84 0.258304009517344
207.36 0.154767912285062
218.88 0.0473080754267001
230.4 0.0396470575789092
241.92 0.0881242277987972
253.44 0.0664803737318902
264.96 0.044903202752602
276.48 0.130212896560065
299.52 0.0838874545221846
311.04 0.0307271112000271
322.56 0.0200283594480335
345.6 0.0287618444223594
368.64 0.0692050739855889
380.16 0.0755336862858456
};
\addlegendentry{Measurement}
\addplot [red, dashed, mark=triangle*, mark size=1.5, mark options={solid}, line width=1pt]
table {%
5.76 66.1030182819925
11.52 34.458564486002
23.04 18.6867766280397
28.8 15.5519618112305
40.32 12.0098375067768
51.84 10.1099880942396
57.6 9.46219133918522
69.12 8.30911817807853
80.64 0.739911405566728
92.16 0.868129090769932
103.68 0.614984598759686
115.2 0.336525657548411
126.72 0.0927475117425449
138.24 0.0936934066229957
149.76 0.203431543909449
161.28 0.192293186260998
172.8 0.106290742065095
184.32 0.279821291338138
195.84 0.245773357162756
207.36 0.157299357205165
218.88 0.0600918722732189
230.4 0.0226863766141125
241.92 0.0715432635720958
253.44 0.0586328944795724
264.96 0.0466752141966742
276.48 0.131731763512127
299.52 0.0904692113144518
311.04 0.0384480182063398
322.56 0.00838371281556105
345.6 0.0272429774702974
368.64 0.0754071140398405
380.16 0.0807231483720554
};
\addlegendentry{Simulation}

\addplot [blue, dash dot, line width=1pt]
table {%
	108.62 0.0001
	108.62 1000
};
\addlegendentry{$\mathrm{B}_{\mathrm{min}}$}

\end{axis}

	\end{small}
\end{tikzpicture}%
		\label{fig_first_case}}
	\hfil
	\subfloat[]{
\begin{tikzpicture}
		\begin{small}

\definecolor{darkgray176}{RGB}{176,176,176}

\begin{axis}[
legend cell align={left},
legend style={fill opacity=0.8, draw opacity=1, text opacity=1, draw=none},
log basis y={10},
tick pos=both,
x grid style={darkgray176},
xlabel={Bandwidth (MHz)},
xmajorgrids,
xmin=-12.96, xmax=398.88,
xtick style={color=black},
xtick={-100,0,100,200,300,400},
xticklabels={
  \(\displaystyle {\ensuremath{-}100}\),
  \(\displaystyle {0}\),
  \(\displaystyle {100}\),
  \(\displaystyle {200}\),
  \(\displaystyle {300}\),
  \(\displaystyle {400}\)
},
y grid style={darkgray176},
ylabel={Estimation error (m)},
ymajorgrids,
ymin=0.0003, ymax=100,
ymode=log,
ytick style={color=black},
ytick={1e-05,0.0001,0.001,0.01,0.1,1,10,100,1000},
yticklabels={
  \(\displaystyle {10^{-5}}\),
  \(\displaystyle {10^{-4}}\),
  \(\displaystyle {10^{-3}}\),
  \(\displaystyle {10^{-2}}\),
  \(\displaystyle {10^{-1}}\),
  \(\displaystyle {10^{0}}\),
  \(\displaystyle {10^{1}}\),
  \(\displaystyle {10^{2}}\),
  \(\displaystyle {10^{3}}\)
}
]
\addplot [black, mark=square*, mark size=1.5, mark options={solid}, line width=1pt]
table {%
5.76 56.0190499251357
11.52 26.6998548605056
23.04 9.14036059996681
28.8 6.5244919917787
40.32 3.0229973782925
51.84 1.00087917611441
57.6 0.357259305178261
69.12 1.14226037490215
80.64 1.41261869236912
92.16 0.683182838641571
103.68 0.102342801723863
115.2 0.199423714409883
126.72 0.64622374280799
138.24 0.320173637098755
149.76 0.00878763026864782
161.28 0.0375378117693401
172.8 0.38877579443354
184.32 0.203853743020063
195.84 0.00842619518815901
207.36 0.0423475571175365
218.88 0.249799468319928
230.4 0.133859290979221
241.92 0.00197101064189642
253.44 0.00842619518805066
264.96 0.15993317365628
276.48 0.0941156057336077
299.52 0.00918562866416028
311.04 0.10917770300822
322.56 0.0671557173345141
345.6 0.0103247788782355
368.64 0.0538656315040038
380.16 0.000578715935811047
};
\addlegendentry{Measurement}
\addplot [red, dashed, mark=triangle*, mark size=1.5, mark options={solid}, line width=1pt]
table {%
5.76 68.9145917832195
11.52 37.3517770859024
23.04 21.7344073680664
28.8 18.6727513094482
40.32 15.0589871137554
51.84 0.913417754124887
57.6 0.58508934798751
69.12 0.214594102273002
80.64 0.781890908867973
92.16 0.906817715675015
103.68 0.259672103508329
115.2 0.0601759621467339
126.72 0.347115243840352
138.24 0.267121584365106
149.76 0.141327053493526
161.28 0.0260014557253472
172.8 0.198266282538339
184.32 0.128651547235514
195.84 0.0876604211873468
207.36 0.0138505201088526
218.88 0.128778119481518
230.4 0.0767569263734078
241.92 0.0593082380821146
253.44 0.00853448577663851
264.96 0.0904267289419609
276.48 0.0510627604343652
299.52 0.00549675187251442
311.04 0.0670108634310118
322.56 0.0365069521437738
345.6 0.00385131267444816
368.64 0.0273937504314024
380.16 0.0252603039067338
};
\addlegendentry{Simulation}

\addplot [blue, dash dot, line width=1pt]
table {%
	54.21 0.0001
	54.21 1000
};
\addlegendentry{$\mathrm{B}_{\mathrm{min}}$}

\end{axis}

	\end{small}
\end{tikzpicture}%
		\label{fig_second_case}}
	\caption{The estimation error compared to the ground truth, for different signal bandwidths. The black solid line shows the results from the measurement and the red dashed line shows the results form the simulation. The blue dash dotted line represents the minimum bandwidth $\mathrm{B}_{\mathrm{min}}$ required to resolve the responses of reference and measurement path. The shown results correspond to \ac{rf} networks with path differences of (a) $R=\SI{2.76}{\metre}$ and (b) $R=\SI{5.53}{\metre}$.}
	\label{fig_4}
\end{figure*}
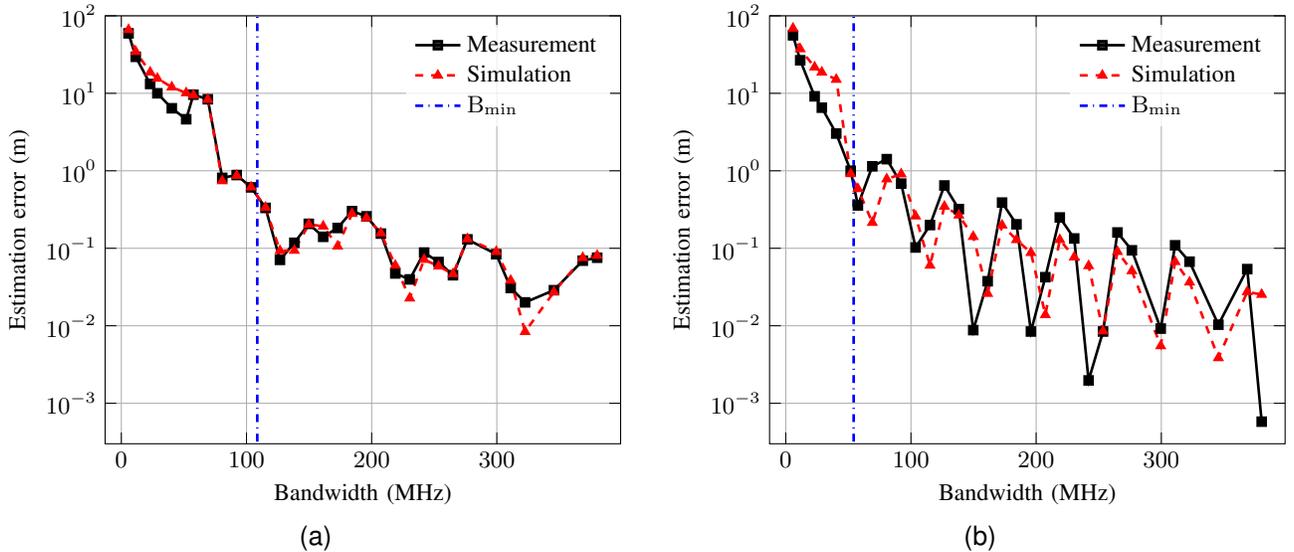

Because of the missing synchronization between TX and RX, a moving average filter, with a window size of $T_{\mathrm{s}}$ is applied on the RX signal to find the start position of the signal within the acquired data.

To determine the ground truth of the range of interest, the S-parameters for all used \ac{rf} paths have been measured with a network analyzer.
\section{Results}\label{sec_4}
Fig.~\ref{fig_3} shows the range profile $\rho[\tau_{n}]$, normalized to the strongest peak for a path difference of $R=\SI{2.76}{\metre}$. Around the two detected peaks, the \ac{czt} results $\rho_{\mathrm{\textsc{czt}}}[\tau_{m,i}]$ are shown in red and orange. The distance between the two peaks is the estimated range $\hat{R}$. Fig.~\ref{fig_4} shows the estimation error \hbox{$|R-\hat{R}|$} from both the measurement setup and the according simulation as function of the bandwidth of the SRS. We also show the minimum bandwidth $\mathrm{B}_{\mathrm{min}}$ which allows to resolve the peaks stemming from the reference and the measurement path
\hbox{$\mathrm{B}_{\mathrm{min}} \geq \mathrm{c}_{0}/R$}. Fig.~\ref{fig_4}(a) shows the results from the measurement configuration with a path length difference of \hbox{$R=\SI{2.76}{\metre}$} and Fig.~\ref{fig_4}(b) for a configuration with \hbox{$R=\SI{5.53}{\metre}$}. For \ac{srs} signals with a bandwidth smaller than $\mathrm{B}_{\mathrm{min}}$, the used peak detection algorithm could not distinguish the peaks in most of the cases. Therefore the next biggest side lobe was wrongly identified as target, resulting in an estimation error in the range of several tens of meters. Starting from a \ac{srs} configuration with $\mathrm{B}_{\mathrm{min}}$ the estimation error is in the order of meters and reduces to the order of centimeters in Fig.~\ref{fig_4}(a) and even to the order of millimeters in Fig.~\ref{fig_4}(b) for bandwidths beyond \SI{200}{\mega\hertz}. The almost periodic variation in the estimation error results from the fact, that for each of the two paths in the setup, according to (\ref{deqn_7}) a sinusoidal oscillation is present over a limited amount of subcarriers. Each results in a $\operatorname{sinc}$ in the range profile. Thus, the sidelobes of one $\operatorname{sinc}$ interfere with the main lobe of the other.
%
%
%
\section{Conclusion}
This paper discussed a \ac{ofdm}-based ranging concept, which adheres to the 5G \ac{nr} \ac{3gpp} standard. At first the basic \ac{ofdm} model was introduced, whereas later-on the standard-related details have been highlighted. The \ac{srs} was introduced as 5G compliant signal and according waveform of choice. The measurement setup and the simulations have been explained and finally the results have been presented in Sec. \ref{sec_4}. Range profiles have been presented together with results of the estimation error for two different measurement setups. We demonstrated, that a ranging estimation error in the order of centimeters is achievable if the \ac{srs} is configured with sufficient bandwidth.
\section*{Acknowledgments}
The financial support by the Austrian Federal Ministry for Digital and Economic Affairs, the National Foundation for Research, Technology and Development and the Christian Doppler Research Association is gratefully acknowledged.
	\begin{acronym}
	\acro{jcas}[JCAS]{joint communication and sensing}
	\acro{jrc}[JRC]{joint radar-communications}
	\acro{ue}[UE]{user equipment}
	\acro{nr}[NR]{New Radio}
	\acro{3gpp}[3GPP]{3rd Generation Partnership Project}
	\acro{ofdm}[OFDM]{orthogonal frequency division multiplexing}
	\acro{ul}[UL]{uplink}
	\acro{srs}[SRS]{sounding reference signal}
	\acro{bs}[BS]{base station}
	\acro{fft}[FFT]{fast Fourier transform}
	\acro{papr}[PAPR]{peak-to-average power ratio}
	\acro{cazac}[CAZAC]{constant-amplitude zero-autocorrelation}
	\acro{rf}[RF]{radio frequency}
	\acro{czt}[CZT]{chirp-Z transform}
	\acro{fpga}[FPGA]{Field Programmable Gate Array}
	\acro{fr2}[FR2]{frequency range 2}
\end{acronym}
\bibliographystyle{IEEEtran}
\bibliography{references}


\end{document}